\documentclass[prb,twocolumn,showpacs,floatfix,amsmath,amssymb]{revtex4}
\usepackage{graphicx}
\usepackage{dcolumn}
\usepackage{bm}
\usepackage{amssymb}
\usepackage{graphicx}
\usepackage{amsmath}
\usepackage{xspace}
\begin{document}

\title{Synthesis, Structure and Magnetic Properties of New Layered Iron-Oxychalcogenide
Na$_{2}$Fe$_{2}$OSe$_{2}$}
\author{J. B. He$^{a}$, D. M. Wang$^{a}$, H. L. Shi$^{b}$, H. X. Yang$^{b}$, J. Q. Li$^{b}$, and G. F. Chen$^{a}$}
\email{genfuchen@ruc.edu.cn}

\affiliation{$^{a}$Department of Physics, Renmin University of
China, Beijing 100872, People's Republic of China}

\affiliation{$^{b}$Beijing National Laboratory for Condensed
Matter Physics, Institute of Physics, Chinese Academy of Sciences,
Beijing 100190, People's Republic of China}

\date{\today}

\begin{abstract}
A new layered iron-oxychalcogenide Na$_{2}$Fe$_{2}$OSe$_{2}$ has
been synthesized and structurally characterized by powder X-ray
diffraction. The structure is formed by alternate stacking of the
newly discovered [Fe$_{2}$OSe$_{2}$]$^{2-}$ blocks and double
layers of Na$^+$. Conductivity study shows that
Na$_{2}$Fe$_{2}$OSe$_{2}$ is a semiconductor with activation
energy of 0.26 eV. Magnetic susceptibility and heat capacity
measurements reveal that an antiferromagnetic phase transition
occurs at T$_{N}$=73 K. A broad maximum of magnetic susceptibility
and a slow decay of the specific heat above T$_{N}$, arise as a
result of two-dimensional short-range spin correlation.

\end{abstract}


\maketitle

Low-dimensional magnetic systems have been a major subject of
theoretical and experimental studies in recent years, due to their
special physical properties and potential applications, such as
high temperature superconductivity (HTSC) \cite{Bednorz, MKWu,
Sheng, Tsuei, Sanna} and colossal magnetoresistance (CMR).
\cite{Helmolt, Asamitsu, Moritomo, Coey} The recent discovery of
superconductivity in layered iron-pnictides has evoked a renewed
interest in these low dimensional 3\emph{d} transition-metal
compounds.\cite{Kamihara, GFChen1, XHChen, GFChen2, Ren, Rotter,
Hsu, Wang, Zhu, XLChen} In this system, the \emph{T$_{c}$} has
become the highest among layered transition-metal-based compounds,
except for the high \emph{T$_{c}$} in cuprates. The full details
of the mechanism for such high \emph{T$_{c}$} superconductors are
not yet completely understood. It is accepted that
antiferromagnetic (AF) spin fluctuations may play some underlying
role in the superconductivity.

In general, the crystal structures of high \emph{T$_{c}$} cuprates
are perovskite related structures, and the structural features are
based on a two-dimensional square lattice sheets of
CuO$_{2}$.\cite{Bednorz, MKWu, Sheng} The new type of
superconducting iron-pnictides are based instead on conducting
layers of edge-sharing FeAs$_{4}$/FeSe$_{4}$
tetrahedra.\cite{Kamihara, Rotter, Hsu, Wang, Zhu, XLChen} In most
of those compounds the CuO$_{2}$ sheets or FeAs$_{4}$/FeSe$_{4}$
tetrahedra layers are interleaved by layers of  rare earth
(alkali- or alkaline-earth) cations blocking electrical
conduction. These parent compounds are commonly AF-Mott-insulators
or AF-semimetals, which can become a superconductor upon electron-
or hole-doping. A lot of efforts have been put into searching for
new materials with similar crystal structure. The transition-metal
oxychalcogenide, La$_{2}$O$_{2}$Fe$_{2}$OSe$_{2}$, containing
anti-CuO$_{2}$ type square planar layer, was originally reported
to be an antiferromagnetic insulator (\emph{T$_{N}$} = 93 K) by
Mayer.\cite{Mayer} In this structure, Fe$^{2+}$ is also
four-coordinated by Se$^{2-}$ (located above or below the center
of [Fe$_{4/2}$O$_{4/4}$]$^{2+}$ square unit), formed an
edge-shared octahedral unit [Fe$_{2}$OSe$_{2}$]$^{2-}$. The more
recent band calculation indicated that
La$_{2}$O$_{2}$Fe$_{2}$OSe$_{2}$ is a Mott insulator due to the
narrowing of Fe \emph{d}-electron bands and the enhanced
correlation effects.\cite{JXZHu} This close structural
relationship to iron-pnictides and cuprates make this system more
intriguing, and it is considered to be a good candidate for high
\emph{T$_{c}$} superconductor. Due to its fascinating structure
and physical properties, a few new compounds,
(Ba,Sr)$_{2}$F$_{2}$Fe$_{2}$OSe$_{2}$, with the same
[Fe$_{2}$OSe$_{2}$]$^{2-}$ unit have been subsequently discovered,
just using the fluorite type [Ba$_{2}$F$_{2}$]$^{2+}$
([Sr$_{2}$F$_{2}$]$^{2+}$) block instead of the fluorite type
[La$_{2}$O$_{2}$]$^{2+}$ block in
La$_{2}$O$_{2}$Fe$_{2}$OSe$_{2}$. These new materials were found
to order antiferromagnetically between 83-107 K and was proposed
to be a rare example of a frustrated AF checkerboard spin
lattice.\cite{Kabbour} We also noticed that a close-packed
structure has been found in titanium oxypnictides
Na$_{2}$Ti$_{2}$O\emph{P}$_{2}$ (\emph{P} = As, Sb),\cite{Ozawa1,
Ozawa2} where the edge-shared [Ti$_{2}$O\emph{P}$_{2}$]$^{2-}$
layers interspersed by double layers of Na$^+$. The occurrence of
a charge-density-wave- (CDW) or spin-density-wave-like (SDW)
instability was reported from the electrical resisitivity and
magnetic susceptibility measurements, which are quite similar to
the parent compounds of high \emph{T$_{c}$} iron-oxypnictides.

Recently, we explored the range of such materials and discovered a
new AF semiconductor Na$_{2}$Fe$_{2}$OSe$_{2}$, which is build by
alternate stacking of [Fe$_{2}$OSe$_{2}$]$^{2-}$ blocks and double
layers of Na$^+$ along c-axis. It is the first example of layered
iron-oxychalcogenide with alkali metal. The temperature dependence
of magnetic susceptibility and heat capacity shows that this
material orders antiferromagnetically around \emph{T$_{N}$} = 73 K
and a two-dimensional (2D) short-range correlation persists up to
temperatures of at least two times T$_{N}$.

Polycrystalline sample of Na$_{2}$Fe$_{2}$OSe$_{2}$ was
synthesized by solid-state reaction method using Na (2N, lump), Se
(5N, powder), Fe$_{2}$O$_{3}$ (3N, powder), and FeSe as starting
materials. FeSe was prepared by reacting Fe powder with Se powder
at 750 $^{0}$C for 20 hours. Stoichiometric amount of Na lump, Se,
FeSe and Fe$_{2}$O$_{3}$ powder were put into an alumina crucible
and then sealed in quartz tube with Ar under the pressure of 0.4
atmosphere. The quartz tube was heated to 500 $^{0}$C slowly and
held there for 20 hours, and then cooled to room temperature
naturally. In order to improve the homogeneity, the resulting
product was then reground, pressed to a pellet, sealed in quartz
tube and reheated at 600 $^{0}$C for a further 20 hours. Black
single crystals could be obtained for annealing the mixture at
elevated temperature of 700 $^{0}$C over 200 hours. Except for
heat-treatment, all manipulations for sample preparation were
performed in glove boxes with a purified argon atmosphere.

The X-ray diffraction (XRD) data of Na$_{2}$Fe$_{2}$OSe$_{2}$ were
collected at room temperature on a Bruker D8 advance using
Cu\emph{K$\alpha$} radiation (2$\theta$ range of 10-80$^{o}$, a
step width of 0.01$^{o}$ and a counting time of 2s). A low
background, airtight specimen holder ring was used to avoid
environmental effect. The XRD data were analyzed by the Rietveld
method with Rietica 1.7.7 analysis package.\cite{Hunter} The
Pseudo-Voigt (Riet asym) function was used as a profile function.
The fixed background was applied by a cubic spline function.
Isotropic atomic displacement parameters, \emph{B}, with the
isotropic Debye-Waller factor represented as
exp((-\emph{B}sin$^{2}$$\theta$)/$\lambda$$^{2}$) were assigned to
all the sites. The resistivity was measured by a standard 4-probe
method. The DC magnetic susceptibility was measured with a
magnetic field of 1 T. The specific heat was measured by a
standard calorimetric relaxation technique. These measurements
were performed down to 2 K in a physical property measurement
system (PPMS) of Quantum Design with the vibrating sample
magenetometer (VSM) option provided.

\begin{table}[h]
\caption{\label{t1} Refined Structural Parameters for
Na$_{2}$Fe$_{2}$OSe$_{2}$ at 300 K (Space Group \emph{I4/mmm} (No.
139); Z = 2; a = 4.107(8) $\AA$, c = 14.641(8) $\AA$, and V =
247.07 $\AA$$^{3}$, agreement factor \emph{R$_{p}$}= 3.52$\%$,
\emph{R$_{wp}$}= 7.43$\%$,  \emph{R$_{exp}$}= 2.10$\%$, S = 3.538.
The occupation (g) of all the sites is unity.)}

\begin{tabular}{p{1.0cm}p{1.0cm}p{1.0cm}p{1.0cm}p{1.6cm}p{1.6cm}}
\hline
\rule{0em}{1em}
atom & site & x & y & z & B($\AA$$^2$)\\
\hline \rule{0em}{1em}
Na & 4e & 0 &  0  &  0.3363(8) & 0.6110(2)\\
\rule{0em}{1em}
Fe & 4c & 0 &  1/2  &  0 & 0.1785(1)\\
\rule{0em}{1em}
Se & 4e & 0 &  0  &  0.1277(7) & 0.5208(1)\\
\rule{0em}{1em}
O  & 2b & 0 &  0  &  1/2 & 0.3316(2)\\

\hline

\end{tabular}
\end{table}

\begin{figure}[h]
\centering
\includegraphics[width=8 cm]{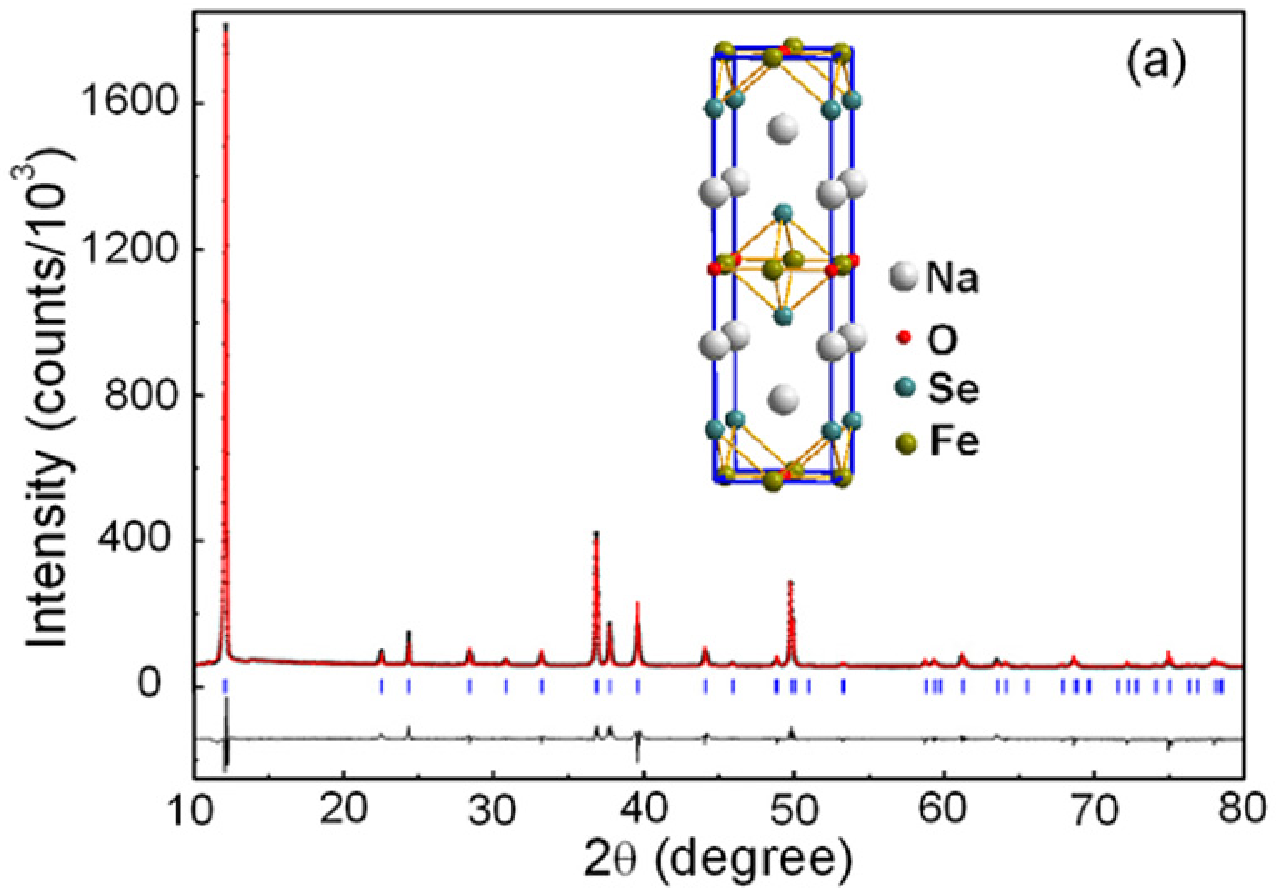}
\hspace{0.5in}
\includegraphics[width=8 cm]{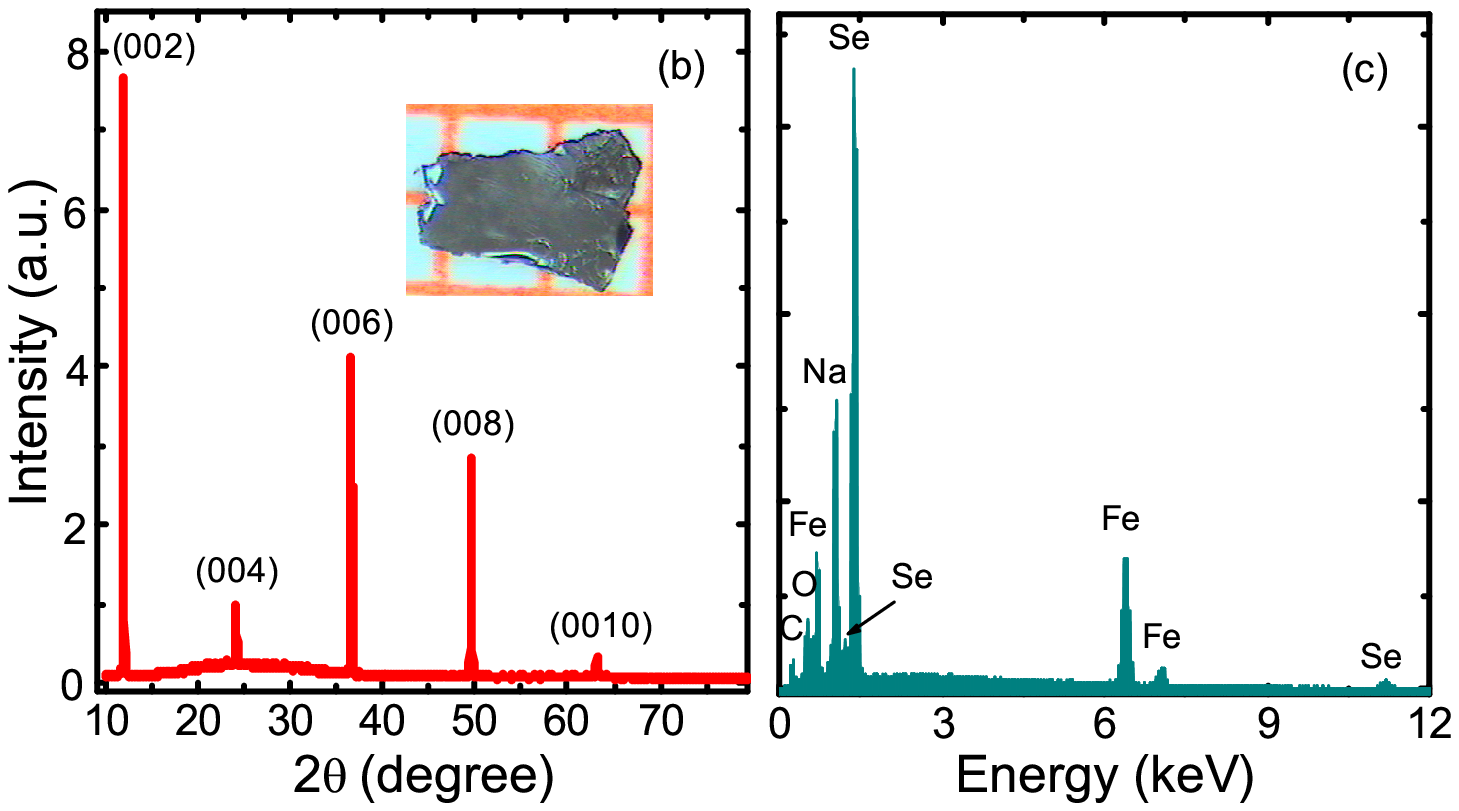}
\caption{(Color online) (a) Powder X-ray diffraction pattern at
room temperature for Na$_{2}$Fe$_{2}$OSe$_{2}$. The solid line
represents the intensities calculated using the Rietveld method.
The bottom curves are the differences between the experimental and
calculated intensities. The vertical lines indicate the Bragg peak
positions of the target compound. The inset shows the crystal
structure of Na$_{2}$Fe$_{2}$OSe$_{2}$. (b) Single crystal X-ray
diffraction spectrum shows only the (00$\ell$) reflections,
indicating that the crystal is cleaved along the a-b plane. Inset:
the picture of single crystal. (c)The EDX spectrum taken on a
piece of single crystal.}
\end{figure}

Figure 1(b) shows the X-ray diffraction pattern for an as-grown
single crystal with the 00$\ell$ ($\ell=2n$) reflections. The
lattice constant c = 14.67 $\AA$ was calculated from the higher
order peaks, comparable to the result from powder diffraction as
described later. In Fig. 1(c) we present the energy dispersive
x-ray microanalysis (EDX) spectrum taken on a piece of single
crystal, which confirmed the presence of the Na, Fe, Se, and O
elements. To determine the crystal structure, high quality powder
XRD data were collected at room temperature, as shown in Fig.
1(a). The crystal structure of
Na$_{2}$Ti$_{2}$OSb$_{2}$\cite{Ozawa1} in the space group
\emph{I4/mmm} was used as a starting model for the final Rietveld
refinement of the Na$_{2}$Fe$_{2}$OSe$_{2}$ structure. The refined
structure parameters are summarized in table 1 and the crystal
structure is illustrated in the inset of Fig. 1(a). The lattice
parameters are determined to be a = 4.107(8) $\AA$ and c =
14.641(8) $\AA$. Na$_{2}$Fe$_{2}$OSe$_{2}$ crystalizes in an
anti-K$_{2}$NiF$_{4}$ type structure, built from alternate
stacking of [Fe$_{2}$OSe$_{2}$]$^{2-}$ blocks and double layers of
Na$^+$ along c-axis. In the [Fe$_{2}$OSe$_{2}$]$^{2-}$ unit,
Fe$^{2+}$ ion is located between oxygen atoms, forming a
square-planar layer, [Fe$_{4/2}$O$_{4/4}$]$^{2+}$ (which is an
anti-configuration to the [CuO$_{4/2}$]$^{2-}$ layer observed in
high \emph{T$_{c}$} cuprates), and two Se$^{2-}$ ions are located
above or below the center of [Fe$_{4/2}$O$_{4/4}$]$^{2+}$ square
unit. In such layered compound, the inter-plane interaction is
expected to be much weaker than the intra-layer one, which is
beneficial to the formation of 2D magnetism and short-range
correlation.

\begin{figure}[t]
\includegraphics[width=8 cm]{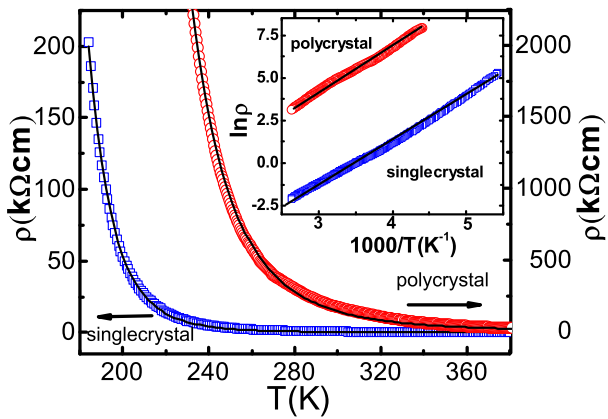}
\caption{(Color online) Temperature dependence of electrical
resistivity for poly- and single-crystal samples of
Na$_{2}$Fe$_{2}$OSe$_{2}$. Inset: The natural logarithm of
electrical resistivity ln$\rho$ plotted against reciprocal of
temperature.}
\end{figure}

Figure 2 shows the temperature dependence of electrical
resistivity measured for poly- and single-crystal samples of
Na$_{2}$Fe$_{2}$OSe$_{2}$. Both of them show an insulating
behavior. The overall resistivity of the polycrystal was nearly
two orders higher than that of single crystal, which might be
ascribed to the apparent grain boundary resistivity. The room
temperature resistivity of Na$_{2}$Fe$_{2}$OSe$_{2}$ single
crystal is about 500 $\Omega$cm, comparable to the value observed
for polycrystalline La$_{2}$O$_{2}$Fe$_{2}$OSe$_{2}$ (100
$\Omega$cm).\cite{Mayer, JXZHu} The electrical resistivity can be
described by an Arrhenius temperature dependence, $\rho$(T) =
$\rho$$_{0}$exp(E$_{a}$/k$_{B}$T), where $\rho$$_{0}$ is a
preexponential factor, E$_{a}$ is the activation energy, and
k$_{B}$ is the Boltzmann constant. The activation energies E$_{a}$
were estimated to be 0.26 eV and 0.24 eV for single- and
poly-crystal, respectively, which are larger than that of
La$_{2}$O$_{2}$Fe$_{2}$OSe$_{2}$ (0.19 eV).\cite{JXZHu} We
attribute this difference to electron hoping between adjacent spin
sites in such a magnetic semiconductor. Our Rietveld refinement
results showed that it has longer Fe$\cdot\cdot\cdot$Fe, Fe---Se,
Fe---O distances for Na$_{2}$Fe$_{2}$OSe$_{2}$, therefore a lesser
integral for hoping between adjacent Fe$^{2+}$ sites was expected
to account for the smaller electrical conductivity.

\begin{figure}[t]
\includegraphics[width=8 cm]{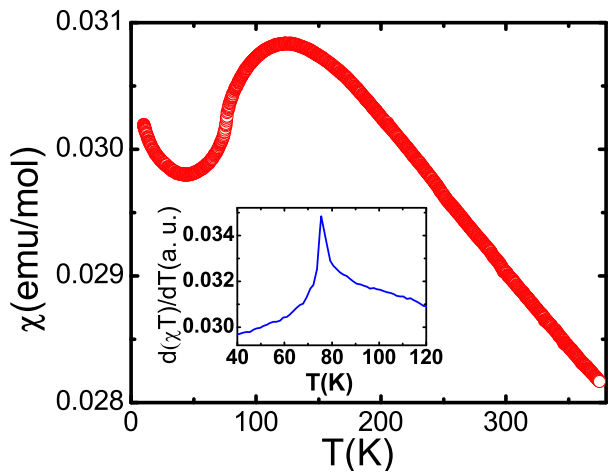}
\caption{(Color online) Temperature dependence of DC magnetic
susceptibility measured at 1 T. The inset shows
\emph{d($\chi$T)/dT}.}
\end{figure}

Figure 3 shows the temperature dependence of DC magnetic
susceptibility measured at 1 T for polycrystalline
Na$_{2}$Fe$_{2}$OSe$_{2}$, which shows a broad maximum around
\emph{T$_{max}$} = 125 K and a clear point of inflexion at
\emph{T$_{N}$} = 75 K. This anomaly is clearly seen in
\emph{d($\chi$T)/dT} (Inset of Fig. 3). The pronounced peak at
\emph{T$_{N}$} indicates the onset of a long-range AF ordering for
\emph{T $<$ T$_{N}$}. Broadening of the susceptibility maximum and
deviation from Curie-Weiss law even at high temperatures, indicate
the strong AF short-range correlations, which is the
characteristic behavior observed in many 2D magnetic systems.
Usually, the ratio of \emph{T$_{N}$/T$_{max}$} is often used to
evaluate the extent of these low-dimensional magnetic
correlations. Applying this simple criterion to our spin system
(\emph{T$_{N}$/T$_{max}$} = 0.58), it could be concluded that
Na$_{2}$Fe$_{2}$OSe$_{2}$ is a typical 2D magnetic system. In
general, in a layered magnetic compound, competition between
intra-layer exchange interactions leads to the possibility of
magnetic frustration effects. Indeed, a systemic studies on
(Ba,Sr)$_{2}$F$_{2}$Fe$_{2}$OSe$_{2}$ showed that these compounds
have a frustrated checkerboard spin lattice.\cite{Kabbour} Similar
effect was recently suggested also to play an important role in
describing the magnetic structure of the isostructural
La$_{2}$O$_{2}$Fe$_{2}$OSe$_{2}$.\cite{David} Hence a shorter
distance between [Fe$_{2}$O] layers in Na$_{2}$Fe$_{2}$OSe$_{2}$
should enhance the frustration effect. A neutron experiment will
provide more useful information to clarify this issue.

\begin{figure}[t]
\includegraphics[width=8 cm]{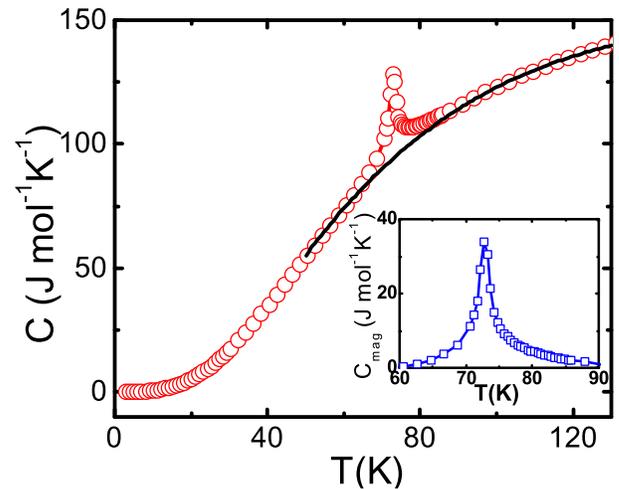}
\caption{(Color online) Heat capacity data measured for
polycrystalline Na$_{2}$Fe$_{2}$OSe$_{2}$. The solid curve
represents the lattice contribution, fitted by a polynomial.
Inset: The estimated magnetic contribution.}
\end{figure}

In order to further clarify the magnetic anomaly, we performed a
heat capacity measurement. The temperature dependence of heat
capacity \emph{C} was plotted in Fig. 4. A clear $\lambda$-type
shape of the \emph{C} peak was observed at \emph{T$_{N}$} = 73 K,
which is due to a long-range antiferromagnetic ordering of the
Fe$^{2+}$ ion. The heat capacity tail due to short-range ordering
of the 2D spin system can be seen more clearly above the
transition temperature. The value of \emph{T$_{N}$} determined
from specific heat is somewhat lower than that one determined from
DC magnetic susceptibility. The low temperature \emph{C(T)} curve
is well fitted by a cubic term $\beta$\emph{T}$^{3}$. From the
value of $\beta$ = 0.64 mJmol$^{-1}$K$^{-4}$, we estimate the
Debye temperature \emph{$\Theta$$_{D}$} = 277 K using the formula
\emph{$\Theta$$_{D}$}=[12\emph{$\pi$$^{4}$NR}/(5$\beta$)]$^{1/3}$.
Because it is impossible to synthesize an analogous non-magnetic
crystal with the same structure to determine the correct lattice
specific heat, we have to estimate the phonon contribution by
fitting a polynomial to the C versus T curve at temperatures well
away from \emph{T$_{N}$} in the data, as shown in the inset of
Fig. 4. By this rough approximation the magnetic contribution to
the entropy over the entire temperature range (\emph{S$_{mag}$} =
2.4 Jmol$^{-1}$K$^{-1}$) seems too much smaller than the
theoretical estimation for a completely ordered Fe$^{2+}$ spin
system (\emph{S} = \emph{R}ln5 = 13.4 Jmol$^{-1}$K$^{-1}$). This
shortfall of entropy implies that the background substraction is
not accurate enough to fully account for the short-range order
entropy developed above \emph{T$_{N}$}. This value is comparable
to that of La$_{2}$O$_{2}$Fe$_{2}$OSe$_{2}$, in which the entropy
release is less than 15$\%$ of \emph{R}ln5.\cite{Fuwa} We noticed
that the features of \emph{C$_{mag}$} and \emph{d($\chi$T)/dT}
(inset of Fig. 3) curves are in good qualitative agreement, due to
the quantity of \emph{d($\chi$T)/dT} being a direct representative
of the magnetic contribution to the heat capacity.

In summary, we have successfully synthesized the single- and
poly-crystals of a new layered iron-oxychalcogenide
Na$_{2}$Fe$_{2}$OSe$_{2}$ , which have Fe$_{2}$O square planar
layers with an anti-CuO$_{2}$-type structure. The electrical
resistivity, magnetic susceptibility and specific heat
measurements indicated that this compound is a magnetic
semiconductor and undergoes a long range AF magnetic transition
below \emph{T$_{N}$} = 73 K. The broadening of the magnetic
susceptibility and the slow decay process of the specific heat
above \emph{T$_{N}$}, indicate an extensive short-range order
above the AF transition point. This behavior might be ascribed to
the frustrated AF exchange interaction in such a 2D checkerboard
spin lattice.

This work was supported by the Natural Science Foundation of China
and the Ministry of Science and Technology of China.

\end{document}